\documentclass[aps,pra,twocolumn]{revtex4-1}

\usepackage{graphicx}
\usepackage{amssymb}
\usepackage{amsmath}
\usepackage{epstopdf}

\begin{document}

	\title{Dynamic diffractive resonant radiation in a linearly chirped nonlinear waveguide array}

	\author{Anuj P. Lara}
	\affiliation{Department of Physics, Indian Institute of Technology Kharagpur, Kharagpur-721302, India}
	\author{Samudra Roy$^{\dagger}$}
	\affiliation{Department of Physics, Indian Institute of Technology Kharagpur, Kharagpur-721302, India}
	\affiliation{Centre of Theoretical Studies, Indian Institute of Technology Kharagpur, Kharagpur-721302, India}
	\email{$^\dagger$samudra.roy@phy.iitkgp.ac.in}

	\begin{abstract}
		 We theoretically and numerically investigate the evolution of discrete soliton in a 1D linearly chirped nonlinear waveguide array (WA). The discrete soliton is self-accelerated inside the transversely chirped WA and emits a \textit{dynamic diffractive resonant radiation} (DifRR). The radiation appears when soliton wave-number matched with linear radiation wave. Unlike uniform WA, the DifRR can be excited even for zero wave-number of input soliton when the  waveguide channels are chirped. The transverse modulation due to chirp conceptually imposes a linear potential which acts as a perturbation to soliton dynamics and leads to a monotonous wave-number shift of the propagating wave. Exploiting perturbative variational analysis we determine the equation of motion of soliton wave-number and use it to establish a modified phase-matching condition which takes into account the soliton wave-number shift and efficiently predicts the  dynamic DifRR. A startling effect like generation of dual DifRR occurs as a result of the interplay between self-accelerated soliton and its initial wave-number. We  exploit the modified phase matching relation to understand this unique phenomenon of dual radiation and find a satisfactory agreement with numerical result in radiation wave-number calculation.

	\end{abstract}
	\maketitle
	\section{Introduction}
		Waveguide Arrays (WAs) since their first inception \cite{1}  has provided a strong platform to study discrete phenomenon which are fundamental in nature. In a WA, a large number (infinite in principle) of single-mode waveguide-channels are placed periodically such that their individual modes overlap and the evolution of an optical field can be represented as a discrete problem. Periodic photonic structures can afford additional control of light, making it possible to explore new physical regimes that are forbidden in homogeneous systems. Discrete diffraction \cite{2}, discrete solitons \cite{3,4} and their interaction with the periodic refractive index lattice are few examples of light management studied in great detail in past years \cite{5}. For discrete soliton, transverse index array is analogous to the continuous temporal counterpart of optical soliton excited in an optical fiber. The discrete nature of spatial soliton introduces additional exciting properties to their characteristics like the Peierls-Nabarro potential \cite{6}, Bloch oscillations \cite{7} and Anderson localization \cite{8}.
Modulation of the periodic structure in uniform homogeneous WA provides additional degrees of freedom in terms of binary WA which offers richer optical property. Binary arrays which are composed of waveguides with different wave numbers allow us to appreciate an optical approach to study relativistic phenomena such as Bloch-Zener Oscillations \cite{9}, Zitterbewegung \cite{10},  Dirac soliton \cite{11}, neutrino oscillations \cite{12}, to name a few.  In addition, introduction of an amplitude and frequency  modulation in the WA have been used to implement beam steering or routing \cite{13,14} and formation of surface solitons \cite{15}. Further extending of these phenomena in the plasmonic regime with beam focusing  in metallic WA \cite{16}, and plasmonic Bloch oscillations in metal-dielectric and graphene arrays \cite{17,18} proves the versatility of this system. 
		In homogeneous WAs, discrete solitons arise due to the stable balance of discrete diffraction and self focusing originating from Kerr nonlinearity. The evolution of electric fields in these WAs are given by the couple mode equations (CMEs) which describe the dynamics of the modes in individual waveguides.
		These equations take into account the self propagation of field in a waveguide, both linear and nonlinear as well as the inter-waveguide interactions that take place through the coupling of the modes by the evanescent electric fields. Although different properties of these discrete solitons have been studied over the years, the phenomenon of these solitons emitting a radiation is a comparatively recent development \cite{19} in this field.
		This radiation, aptly named \textit{diffractive resonant radiation} (DifRR)\cite{19}, is emitted by special soliton propagating in a uniform WA. Such radiation is the spatial (or wavenumber) analogue to dispersive radiation emitted from a ultrashort pulse in an optical fiber \cite{20}. The presence of higher order dispersion in fibers leads to a phase-matching (PM) situation which allows the soliton to transfer energy to the linear dispersive waves at specific frequencies. Similar to its temporal counter part, static DifRR having a specific wave-number is emitted  when soliton wave-number matches with the linear wave propagating in a WA. However, the Brillouin boundary due to the 1D lattice created by the periodic arrangement of waveguides limits the possible wavenubers to lie within $-\pi$ and $\pi$. Any electric field going beyond this boundary undergoes a $2\pi$ shift and emerge from the other side of the boundary. This unusual effect is termed as anomalous recoil \cite{19}. We will see in the later sections that an initial wavenumber is required to generate the DifRR \cite{19} where as it can be controlled by some other parameters like soliton power and coupling coefficient. 
			
		In this work we mainly investigate the dynamics of DifRR emitted by a discrete soliton in a geometrically modified non-uniform WA. A modification in WA provides a versatile platform in controlling light where propagating optical field experiences perturbation. An instability can be introduced to perturb the optical field by providing an external irregularities in the WA either by modifying the refractive index or waveguide arrangement.  A constant difference of propagation vector in adjacent channel arises due to the transverse index gradient shows exciting dynamics even in linear domain where optical analogy of Bloch oscillations is identified \cite{21,22}. In another scheme, the coupling coefficient of the WA are randomly varied by changing their relative positions of waveguide channels which offers Anderson localization \cite{23}. Inspired by these works we make an attempt to understand the optical field dynamics inside a linearly chirped 1D WA which is less explored in the context of discrete soliton propagation. In a linearly chirped WA the separation between adjacent waveguides increases (or decreases) with an uniform rate called \textit{chirp parameter} which leads to a variation in the coupling coefficient. DifRR is found to be an inevitable phenomenon in chirped WA where solion moves with a self-accelerated mode. The chirp conceptually acts as a linear potential that perturb the soliton propagation and leads to dynamic DifRR where radiation wave-number shifts along propagation distance. The soliton dynamics under linear potential is theoretically estimated exploring perturbative analysis based on variational theory. Exploring these results we establish a modified PM expression which predicts the dynamic DifRR accurately for chirped WA. Further we extend our investigation to DifRR formation under non-zero initial soliton wave-number ($k_0 \neq0$).  The interplay between the chirp parameter and $k_0$  opens up new operational domain previously not possible. Here we find an unique case for $k_0>0$ where dual DifRR appears which was never observed before. Based on theoretical analysis we try to explain the intriguing effect of dual DifRR and the agreement  between numerical and analytical result is satisfactory.   
 	\section{Theory}
\begin{figure}[h]
			\begin{center}
				\includegraphics[width=1.05\linewidth]{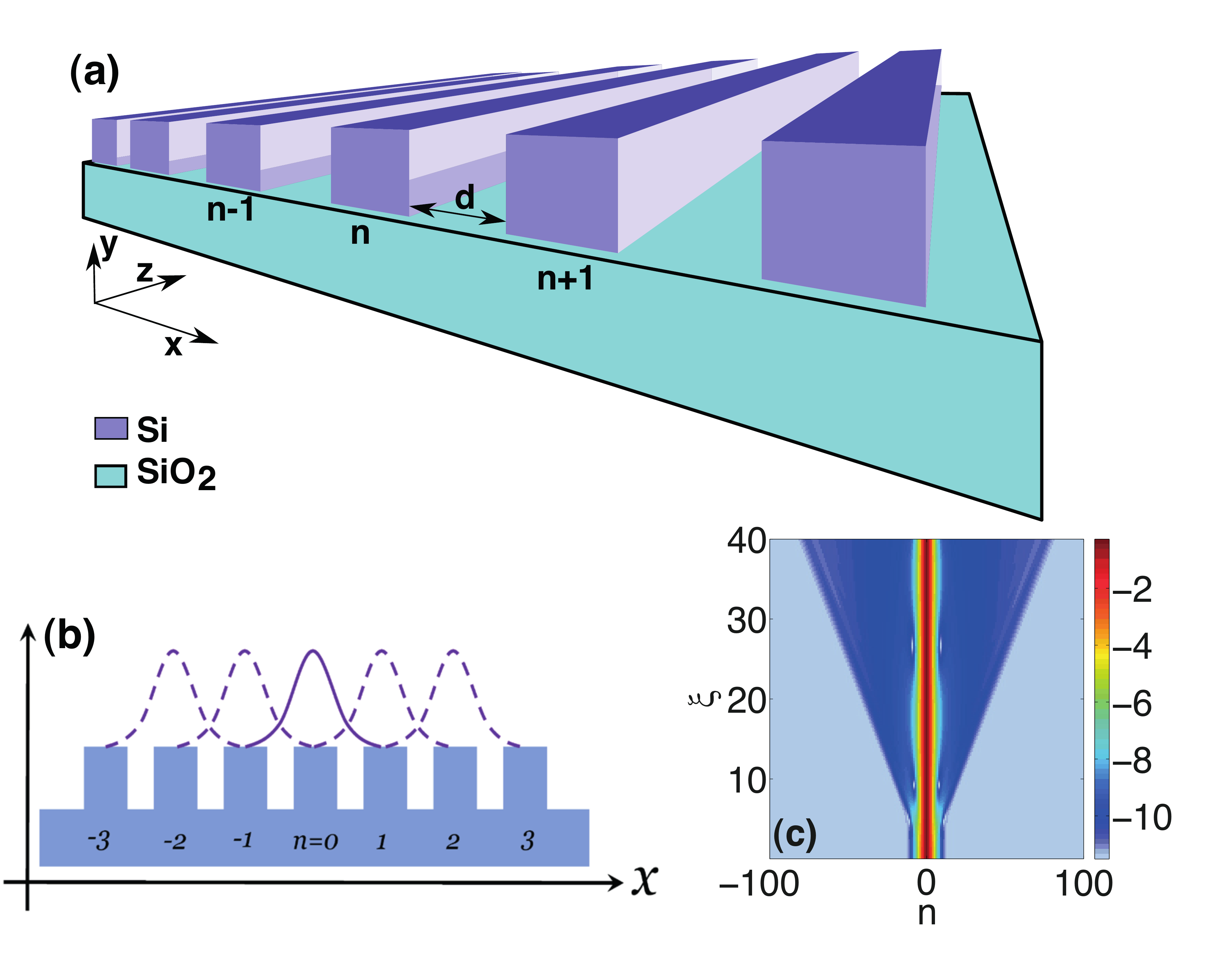}
				\caption{(a) A uniform WA having inter-wavguide separation \textit{d}. (b) Schematic representation of nearest neighbor evanescent mode coupling. (c) Formation of discrete soliton in nonlinear WA.}
				\label{fig:WA}
			\end{center}
		\end{figure}

	\noindent A semi-infinite array of identical periodic nonlinear waveguides with no losses is considered as ideal WAs. For continuous wave excitation in such WA, the evolution of mode amplitude in the $n^{th}$ waveguide with nearest-neighbor evanescent coupling is described  by the discrete nonlinear Schr\"{o}dinger equation (DNLSE)\cite{24},
		\begin{align}
			i \frac{dE_n}{dz} + C_{(n)}^{(n+1)}E_{(n+1)} + C_{(n)}^{(n-1)}E_{(n-1)} +\gamma |E_n|^2 E_n = 0.
			\label{eq:gen_CME}
		\end{align}
	$E_n$ is the electric field amplitude of the $n^{th}$ waveguide and $n=\{1,2,....N \}$, where $N$ is the total no of waveguide.  Here, $C_{(n)}^{(n+1)}$ and $C_{(n)}^{(n-1)}$ are respectively, the coupling coefficients of the $(n+1)^{th}$ and $(n-1)^{th}$ waveguides to the $n^{th}$ waveguide in the unit of 1/m. $\gamma=\omega_0 n_2/(cA_{eff})$ is the nonlinear coefficient of a single waveguide in the unit of 1/Wm where $n_2$ is the Kerr coefficient and $A_{eff}$ is the effective area of the modes. In Fig.(\ref{fig:WA}a) we  represent model of a uniform WA having equal separation between the two consecutive waveuguide channels. The coupling coefficients which are a function of separation, become identical throughout the WA ($C^{(n+1)}_{(n)}=C^{(n-1)}_{(n)}=C$). The nearest neighbor evanescent mode-coupling is schematically illustrated in Fig.(\ref{fig:WA}b) where the sketch of refractive index in lattice is shown.  At low powers the nonlinear term can be neglected ($\gamma=0$) and Eq.(\ref{eq:gen_CME}) can be analytically integrable. A single waveguide excitation leads to a solution $E_n(z)=E_n(0)i^nJ_n(2Cz)$ exhibiting discrete diffraction \cite{4}, where $J_n$ is the Bessel function of order $n$. Physically, the discrete diffraction is originated due to the varying $z$-dependent phase shift for different transverse wavevector components. The discrete diffraction can be restricted by the focusing nonlinearity of the system and one can intuitively understand the soliton formation as a balance between Kerr nonlinearity and diffraction. In Fig.(\ref{fig:WA}c) we demonstrate a discrete soliton that is originated in the  uniform nonlinear WA. For uniform WA (Fig.\ref{fig:WA}a) a useful normalized form of the DNLSE can be realized by making the following transformations $E_n \rightarrow \sqrt{P_0} \psi_n$, $\gamma P_0 z \rightarrow \xi$ and $C/(\gamma P_0) \rightarrow c$, 
		\begin{align}
		i \frac{d\psi_n}{d \xi} + c [\psi_{n+1} + a_{n-1}] + |\psi_n|^2 \psi_n =0,
		\label{eqDNLSE}
		\end{align}
where $P_0$ is the peak power of the associated beam in the units of Watt. Note, the total power $P=\sum\limits_{n}|\psi_n|^2$ and Hamiltonian $H=\sum\limits_{n}[c|\psi_n-\psi_{n-1}|^2-\frac{\gamma}{2}|\psi_n|^4]$ remain conserved during propagation \cite{24}. For a stationary discrete plane wave solution $\psi_n(\xi) = \psi_0 \exp[i(nk_xd + \beta \xi)]$ of Eq.\eqref{eqDNLSE}, one can obtain the dispersion relation between $\beta$ and $k_x$ as \cite{25},
		\begin{align}
			\beta(\kappa) = 2 c \cos(\kappa)+|\psi_0|^2,
			\label{eq:disp}
		\end{align}
where, $d$ is the separation between two adjacent waveguide, $k_x$  is the transverse wave-vector, and $\kappa = k_xd$ represent the phase difference between adjacent waveguides. Note, during propagation the transverse component ($\kappa$) gains a phase $\phi_t=\beta(\kappa)\xi$ which leads to the transverse shift $\Delta n=\partial \phi_t/\partial \kappa$ of the propagating beam \cite{26}. 
The angle $\theta$ of beam propagation follows $\tan \theta=\Delta n/\xi$. Hence $\kappa$ governs the propagation direction as $\theta= \tan^{-1}(\partial \beta(\kappa)/\partial \kappa)=\tan^{-1}(-2c\sin(\kappa))$ \cite{27}. The Taylor expansion of $\beta (\kappa)$ about the incident wavenumber ($\kappa_0$) gives us an expanded diffraction relation,
		\begin{align}
			\beta(\kappa) = \beta(\kappa_0) + \sum_{m\geq 1} \frac{D_m}{m!}\Delta\kappa^m,
		\end{align}
where $D_m \equiv \left(d^m \beta/d \kappa^m\right)|_{\kappa_0}$ and $\Delta\kappa = \kappa - \kappa_0$. The Fourier transformation to change the domain $\kappa \rightarrow n$ is done by replacing $\Delta \kappa \equiv -i\partial_n$ where $n$ is defined as a continuous variable of an amplitude function $\Psi(n,\xi) = \psi_{n,\xi} \exp(-i \kappa_0 n)$ \cite{2,5}. 
		
\noindent  Defining $n$ as continuous variable, which is justfied  as solitons extend for several waveguides, we have an approximated standard nonlinear Schr\"{o}dinger equation (NLSE)\cite{19}
		\begin{align}
			\left[i \partial_\xi - \frac{D_2}{2}\partial_n^2 + \sum_{m\geq 3}\frac{D_m}{m!}(-i\partial_n)^m  + |\Psi(n,\xi)|^2 \right]\Psi(n,\xi)=0.
			\label{eq:NLSE}
		\end{align}
		The first and second terms of the Taylor expansion are eliminated by introducing a phase evolution substitution $\Psi(n,\xi)\rightarrow\Psi(n,\xi) \exp[i \beta(\kappa_0)\xi]$ and using the concept of co-moving frame $n \rightarrow n + D_1 \xi$. For $D_{m\geq 3}=0$, Eq.(\ref{eq:NLSE}) has a soliton solution given by,
		\begin{align}
		\Psi_{sol} = \Psi_0 sech\left(\frac{n \Psi_0}{\sqrt{|D_2|}}\right) \exp(ik_{sol} \xi),
		\label{eq:sol}
		\end{align}
		where $k_{sol} \equiv \Psi_0^2/2$ is the longitudinal wave number for spatial soliton. Note that, for bright soliton solution we have the condition $|\kappa_0|<\pi/2$. The  plane wave solution  $\exp\left[i(k_{lin}\xi + \Delta \kappa n)\right]$ of the linearized Eq.(\ref{eq:NLSE}) gives us the dispersion relation,
		\begin{align}
			k_{lin}(\Delta \kappa) =\beta(\kappa)-\beta(\kappa_0)-D_1\Delta \kappa.
			\label{eq:ksol}
		\end{align}		
A soliton of the form given by Eq.(\ref{eq:sol}) transfers energy to the linear wave and generates a radiation when $k_{sol} = k_{lin}(\Delta \kappa)$ is satisfied. This is the required phase matching (PM) condition for the diffractive resonant radiation (DifRR) as predicted in the seminal paper \cite{19}.
		
	\subsection{Generation of diffractive resonant radiation in uniform waveguide array}
	
\begin{figure}[h]
		\begin{center}
			\includegraphics[width=1.01\linewidth]{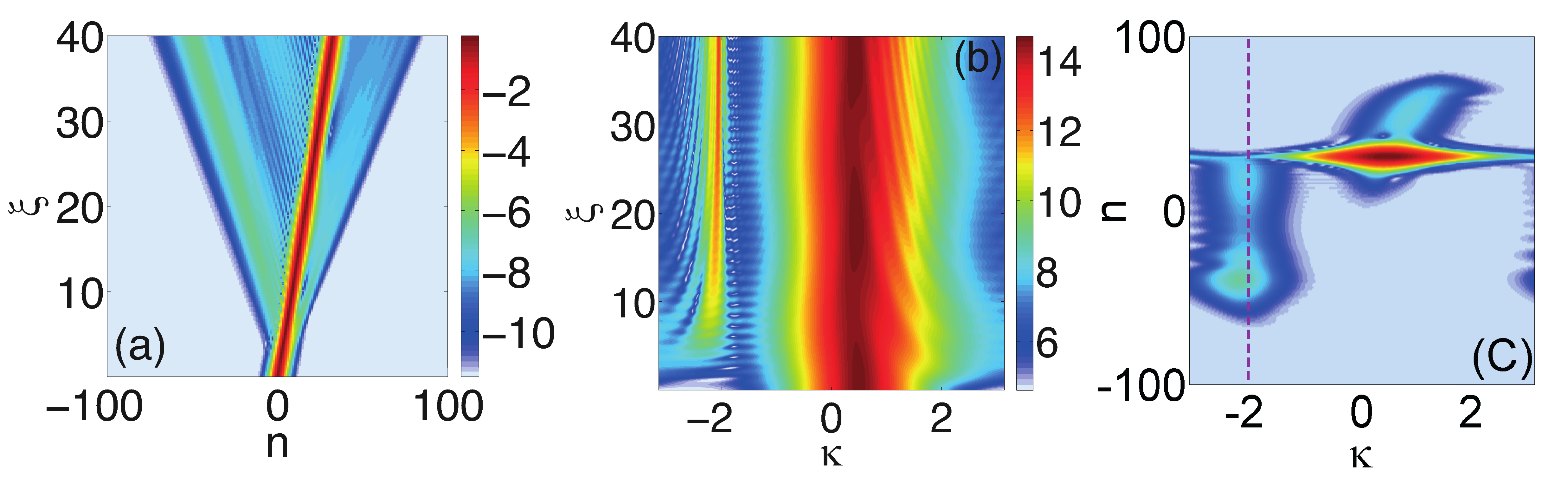}
			
			\caption{ Discrete soliton propagation in (a) $n$-space  and (b) $\kappa$-space for  $\psi_0=0.8$ and $\kappa_0=0.5$. (c) XFROG at $\xi=40$ where diffractive radiation is evident. The vertical dotted line indicate the location of $\kappa_{RR}$ which is obtain by solving Eq.(\ref{PMC}).} 
			\label{Fig2}
		\end{center}
	\end{figure}

\begin{figure}[h]
			\begin{center}
				\includegraphics[width=1.0\linewidth]{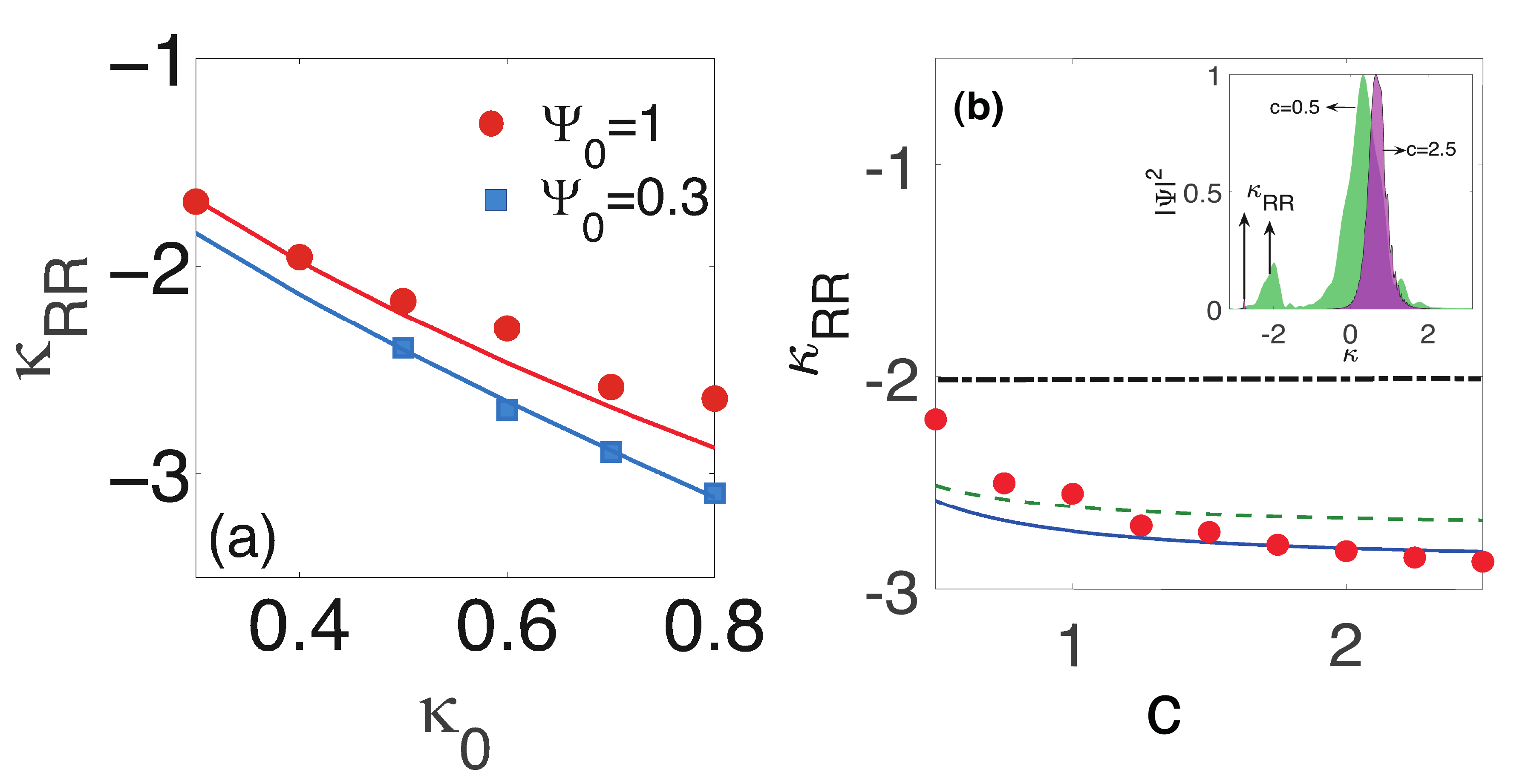}
				\caption{ (a) Location of DifRR ($\kappa_{RR}$) as a function of input soliton wave-number $\kappa_0$ for two different amplitudes. Numerically obtained $\kappa_{RR}$ are presented by solid dots and squares. The solid lines based on the solution of Eq.(\ref{PMC}) theoretically predict $\kappa_{RR}$.  (b) Variation of $\kappa_{RR}$ as a function of coupling constant ($c$). The solid dots are numerical data where solid line represents PM solution of Eq.(\ref{PMC}). The dashed line corresponds to the closed form expression shown in Eq.\eqref{eq:rpmc}. The horizontal dot-dashed line appears when we neglect $\Delta$ in Eq.\eqref{eq:rpmc}. In the inset we show the formation of DifRR for two different coupling constants.   } 
				\label{Fig3}	
			\end{center}
		\end{figure}

\noindent The generation of DifRR requires the soliton to have an initial wavenumber as per the phase matching equation. The PM condition $k_{sol} = k_{lin}(\Delta \kappa)$  leads to a transcendental equation,
		\begin{align}
			\left[\cos(\kappa)-\cos(\kappa_0) + \sin(\kappa_0)\Delta \kappa\right] = \widetilde{\Psi}_0^2,
			\label{PMC}
		\end{align}
		where, $\widetilde{\Psi}_0=\Psi_0/2\sqrt{c}$. The solution of this relation  gives the wavenumber of the generated DifRR ($\kappa_{RR}=\kappa_0+\Delta \kappa$) as a function of initial soliton wavenumber $\kappa_0$. Considering the contribution of the right-hand side is small one can have an approximate solution of Eq.(\ref{PMC}) , 
           \begin{align}
			\kappa_{RR} \approx \kappa_0 + 3 \cot(\kappa_0)\left[1-\frac{1}{4}(1-\sqrt{1+\Delta^2}) \right],
			\label{eq:rpmc}
		\end{align}
		where, $\Delta=\frac{4 \widetilde{\Psi}_0}{3}\frac{\tan\kappa_0}{\sqrt{\cos\kappa_0}}$. The approximated solution is valid under certain range of parameters and consistant with the result given in  \cite{19} if we neglect $\Delta$. In Fig.(\ref{Fig2}) we demonstrate the dynamics of soliton and the formation of DifRR in a uniform WA. The evolution of the input beam with the form $\Psi_{sol}=\Psi_0sech(n\Psi_0/\sqrt{|D_2|})e^{ik_{sol}\xi}$ is shown is in Fig.(\ref{Fig2}a). The soliton emits radiation around $\xi=5$ as demonstrated in the Fourier spectrum of $\Psi(n)$ in Fig.(\ref{Fig2}b). The cross-correlation frequency-resolved optical grating (XFROG) diagram or spectrogram is shown in Fig.(\ref{Fig2}c) where the location of the DifRR ($\kappa_{RR}$) is indicated by vertical dotted line. XFROG is a wellknown technique through which we can plot the wavenumber and its spacial counterpart together. Mathematically it is defined as $\mathcal{S}(n,\kappa,\xi)=|\int_{\infty}^{\infty} \Psi(n^\prime,\xi)\Psi_{ref}(n-n^\prime)e^{i\kappa n^\prime}dn^\prime|^2$
where $\Psi_{ref}$ is the reference window function normally taken as the input. Eq.(\ref{PMC}) is exploited to estimate the location of $\kappa_{RR}$ in $\kappa$-space.  Note, the limits of the $\kappa$ domain lies within the first Brillouin zone ($-\pi < \kappa < \pi$) and if any part of the soliton or DifRR crosses this limit an additional wavenumber of $-2 \pi$  gets added. The Brillouin boundary appears due to the 1D lattice formed by the periodic arrangement of waveguides. This confines the value of the wavenumber to this limit, and the phenomenon is termed as \textit{anomalous recoil} \cite{19,28}. 
From the phase matching equation Eq.\eqref{PMC}, it is evident that DifRR can be tunable under various parameters like initial soliton wave number or momentum ($\kappa_0$), coupling coefficient ($c$) and beam amplitude ($\Psi_0$). In the previous studies \cite{19,28} the dominant role of input wave number ($\kappa_0$) is  mainly investigated in the context of  DifRR formation. Approximate closed form expression $\kappa_{RR}=\kappa_0+3/\tan\kappa_0$ are proposed to deduce the wave number of DifRR. In this work, however, we try to generalize the study by capturing the role of other two parameters $\Psi_0$ and $c$ in the evolution of DifRR. In Fig.(\ref{Fig3}a) we plot the DifRR wave-number as a function of $\kappa_0$  for two different beam amplitude. We find there is a difference in $\kappa_{RR}$ values when we change beam amplitude. The full PM expression Eq.(\ref{PMC}) (solid lines) nicely predicts the DifRR wave-number in both cases. Next we examine the role of coupling coefficient in DifRR generation. Note, coupling coefficient can be easily varied by changing the separation between waveguide channels.  In Fig.(\ref{Fig3}b) we illustrate the variation of DifRR wave-number ($\kappa_{RR}$) with coupling coefficient $c$. The solid dots represent the the values of $\kappa_{RR}$ which are obtained numerically by solving Eq.\eqref{eqDNLSE} where as the solid line corresponds to PM solution of Eq.(\ref{PMC}). The dashed line represents the approximated closed  expression derived in Eq.\eqref{eq:rpmc}. For a comparison we also plot (horizontal dot-dashed line) the closed form expression $\kappa_{RR}=\kappa_0+3/\tan\kappa_0$  derived in \cite{19}. In the inset we show the field distribution $|\Psi_0|^2$ in $k$-space for two different $c$ values where the shift of the radiation is evident. It is also noticed that a stronger, but wide radiation emerges for low values of coupling coefficient $c$ where as a sharp but weak radiation appears when $c$ is comparatively large. We find for low coupling coefficients, numerically it is tricky to determine the exact value of $\kappa_{RR}$ as the radiation spreads over a region. This anomaly in measurement  causes a slight deviation in numerical and analytical result specially for low $c$ values.

\section{Discrete soliton in Chirped Waveguide Array}
  \noindent  The propagation dynamics of discrete soliton becomes more intriguing and practically useful if some non-uniformity is introduced in the WA. Depending on the application, few standard strategies are implemented to bring non-uniformity in WAs like, by changing  the waveguide width \cite{8} or changing the separation between adjacent waveguide \cite{23}. Such WAs are used to describe the Anderson localization in nonlinear optics. In another scheme, optical Bloch oscillations can be realized in WAs with linear refractive index modulation in transverse direction \cite{21}. The linear refractive index variation is mathematically adjusted by incorporating a linear potential term in NLSE. In this work, we have introduced a chirped-WA where the separation between adjacent waveguide changes linearly. The coordinate of $n^{th}$ waveguide is defined as, $x_n=nd_0+\frac{\delta}{2}n(n-1)$, where $d_0$ is the separation between central ($n=0$) to first ($n=1$) waveguide and $\delta$ defines increment of waveguide separation in real unit. We introduce a normalized chirp parameter defined by $g_c=\delta/d_0$ that denotes the strength of chirping. For this system, the propagation constant remains same for all waveguide while the coupling coefficient ($c$) varies along transverse distance. A linearly chirped WA is conceptually realized by a linear potential \cite{29}. Exploiting this concept we may configure a perturbed NLSE as, $[i \partial_\xi + \frac{1}{2}\partial_n^2  + |\Psi(n,\xi)|^2 ]\Psi(n,\xi)= \frac{i}{2} \epsilon$, where $\epsilon=i\chi n \Psi$ accounts for the linear potential term as a perturbation and $\chi$ is related to the potential strength. A standard perturbative variational analysis \cite{30} with a regular ansatz, $\Psi=\eta sech[\eta(n-n_0)]e^{i[\phi-\kappa(n-n_0)]}$ can be exploited to estimate the evolution of the soliton wave number ($\kappa$) and position ($n_0$) under linear potential. The variational treatment ensures the conservation of total energy $\frac{\partial \mathcal{E}}{\partial \xi}=0$, $ (\mathcal{E}=\int|\Psi|^2 dn)$ and leads to the equation of motions,  $\frac{\partial\kappa}{\partial \xi}=-\chi$ and $\frac{\partial n_0}{\partial \xi}=\kappa$. While propagating through a uniform WA ($\epsilon=0$), the soliton maintains its wavenumber. However, an evolution in the wavenumber is implemented when soliton propagates under perturbation like a linear potential appearing transversely along the $n$ coordinate which take into account the chirping. The variational result predicts that  the soliton wavenumber starting at $\kappa_0$ experiences a continuous linear shift  $\kappa(\xi) = \kappa_0 - \chi \xi$. The corresponding evolution in position of soliton is $n_0(\xi) = n_0(0) - \frac{1}{2}\chi\xi^2$ when starting with $\kappa_0 = 0$. A similar evolution of soliton is observed in a WA with quasiperiodic lattice arrangement \cite{31}, which piques the idea of DifRR generation in such systems.

\subsection{ Waveguide design}
\begin{figure}[h]
			\begin{center}
				\includegraphics[width=1.35\linewidth]{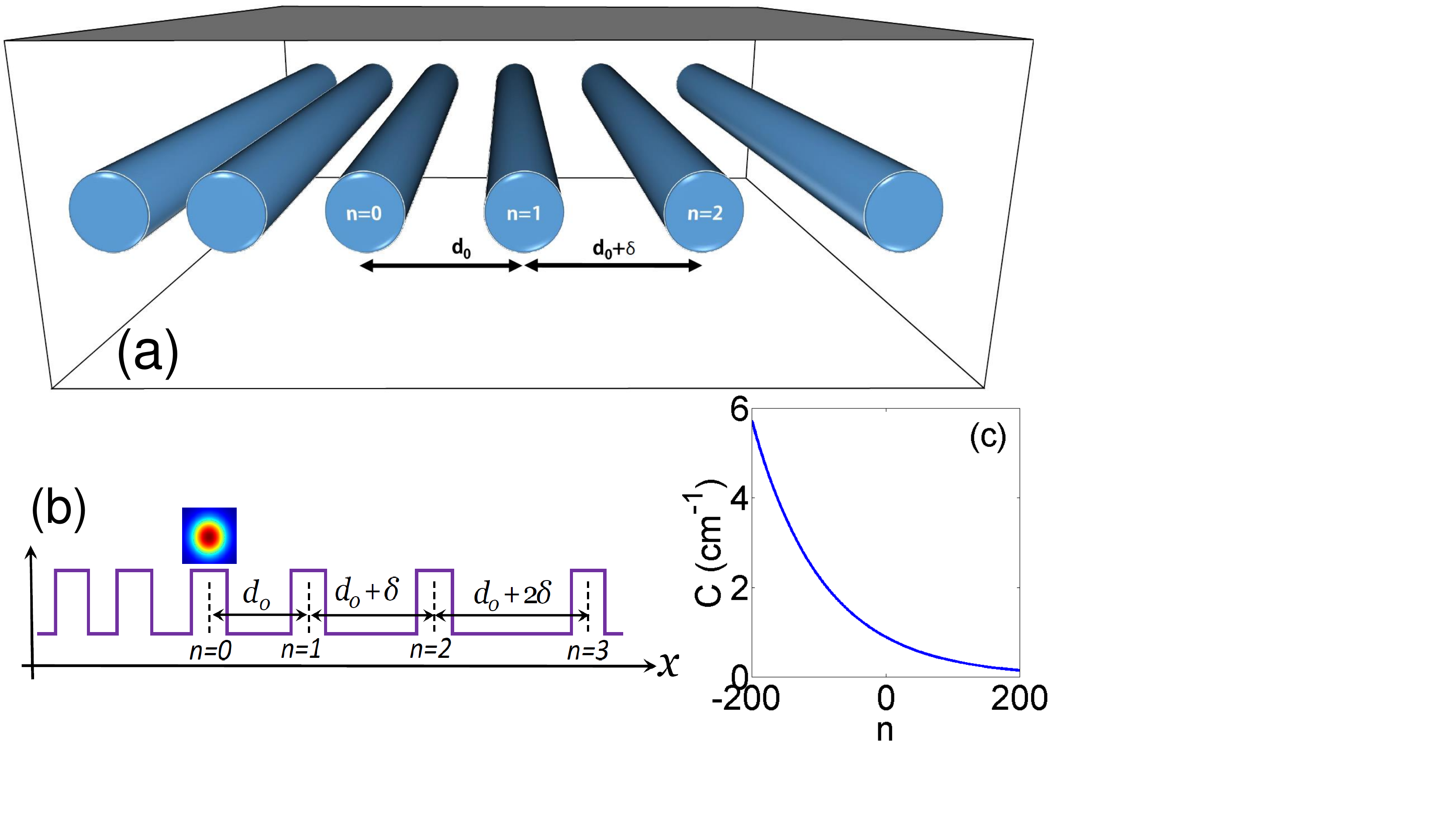}
				\caption{(a) Pictorial model of the proposed WA where the cylindrical channels are placed non-uniformly with increasing separation. (b) Schematic representation of waveguide arrangement where central channel ($n=0$) is illuminated by input electric field. (c) Spacial variation of coupling coefficient in real unit. }
				\label{fig:C_WA}
			\end{center}
		\end{figure}

Befor going to a detailed analysis of soliton dynamics it is important to define a physically realizable waveguide structure that supports DifRR. Strategically, a chirped WA can be formed by modulating either the refractive index of waveguide channels or by their relative separation. Modulation of the refractive index introduces a position dependent propagation vector, while modification of the waveguide separation results in a position dependent coupling coefficient. The facility of fs laser based writing in transparent bulk medium \cite{32,33}
allows us to design a WA of cores suspended in its cladding as modeled in Fig.(\ref{fig:C_WA}a). We propose  GeO$_2$ doped silica cores suspended in a silica cladding, to have an equivalent refractive index difference $\Delta n$, between the core and cladding. At operating wavelength $\lambda_0=1.55$ $\mu$m, the core and cladding refractive indices are $n_1 \approx$1.4477 and $n_2\approx$1.4446, respectively. We consider the radius of the cylindrical core $a=5 \mu$m. For the given geometry of the WA the nonlinear coefficient is calculated as, $\gamma=0.79$ W$^{-1}$km$^{-1}$.   As schematically shown in Fig.(\ref{fig:C_WA}b), a chirped WA is designed by taking an initial separation $d_0$= 20 $\mu$m between the central reference waveguide ($n=0$) and $n=1$ waveguide, then apply a progressive change in separation ($\delta$) in the nm range to keep the resulting perturbation small. Since the separation between waveguides is a function of position, we calculate the coupling coefficients as a function of the respective separation ($d_n$) using \cite{34}
		\begin{align}
			C(d_n) = \frac{\lambda_0}{2\pi n_1} \frac{U^2}{a^2 V^2} \frac{K_0(Wd_n/a)}{K_1^2(W)}.
			\label{eq:coup_calc}
		\end{align}
		
			
\noindent Here $d_n=d_0(1+ng_c)$, $\lambda_0$ is the wavelength in free space (1.55 $\mu$m in this case), $n_1$ and $n_2$ being the core and cladding refractive indices respectively, $a$ the core radius, and $K_v$ are the modified Bessel functions of the second kind of order $v$.
$U$ and $V$ are the mode parameters that satisfy $U^2 + W^2 = V^2$, where V parameter defined by $V=\frac{2 \pi a}{\lambda_0} \sqrt{n_1^2 - n_2^2}$. $U$ is given approximately as, $U \cong 2.405 e^{-(1-\nu/2)/V} $, with $\nu=1-(n_2/n_1)^2$ \cite{35}. In Fig.(\ref{fig:C_WA}c) we depict the variation of the coupling coefficient ($C$) in real unit for the proposed chirped WA.

		\subsection{Generation of dynamic DifRR in a linearly chirped waveguide array}
		
\noindent In this section we numerically investigate the evolution of a discrete soliton Eq.(\ref{eq:sol}) in a linearly chirped WA and formation of dynamics DifRR.  We can construct a normalized set of DNLSE from Eq.(\ref{eq:gen_CME}) by taking the  transformations $\eta_n^{n\pm1} \rightarrow C_{(n)}^{(n \pm 1)}/C_0$, $\xi \rightarrow C_0 z$, $E_n \rightarrow \sqrt{P_0}a_n$, and $\psi_n \rightarrow \psi_0 a_n$, where $\psi_0 \rightarrow \sqrt{\gamma P_0 /C_0}$,
		\begin{align}
			i d_{\xi} \psi_n\left(\xi\right) + \eta_{(n)}^{(n+1)} \psi_{n+1}\left(\xi\right) + \eta_{(n)}^{(n-1)}\psi_{n-1}\left(\xi\right) \nonumber \\
			+ |\psi_n\left(\xi\right)|^2 \psi_n\left(\xi\right) = 0.
			\label{eq:norm_CME}
		\end{align}
Here $C_0$ is the coupling coefficient between central ($n=0$) and  1$^{st}$ ($n=1$) waveguide channel whose value is calculated to be $\approx$ 1 cm$^{-1}$.  Equation \eqref{eq:norm_CME} mathematically describes the soliton evolution in the chirped WA.  In a preliminary analysis, we numerically investigate the dynamics of soliton inside the proposed  WA which is schematically shown in Fig.(\ref{fig:C_WA}a).	We launch the soliton in a photonic system where the coupling coefficient varies along the direction transverse to the propagation direction. The value of $g_c$  determines the rate at which the coupling coefficient changes. It is apparent from Eq.(\ref{eq:coup_calc}) that the value of the coupling coefficient (C) will decrease as the separation increases owing to the decaying nature of modified Bessel function of second kind $K_{j}(x)$. The value of $\delta$ is considered small ($\sim$ nm) compare to the separation ($\sim$ $\mu$m) between waveguides. The power scale $P_0 \sim$ 125 kW makes the scaling factor $\psi_0$ to unity and length scale become $\sim$ 1 cm. 
		\begin{figure}[h]
			\begin{center}
				\includegraphics[width=1.05\linewidth]{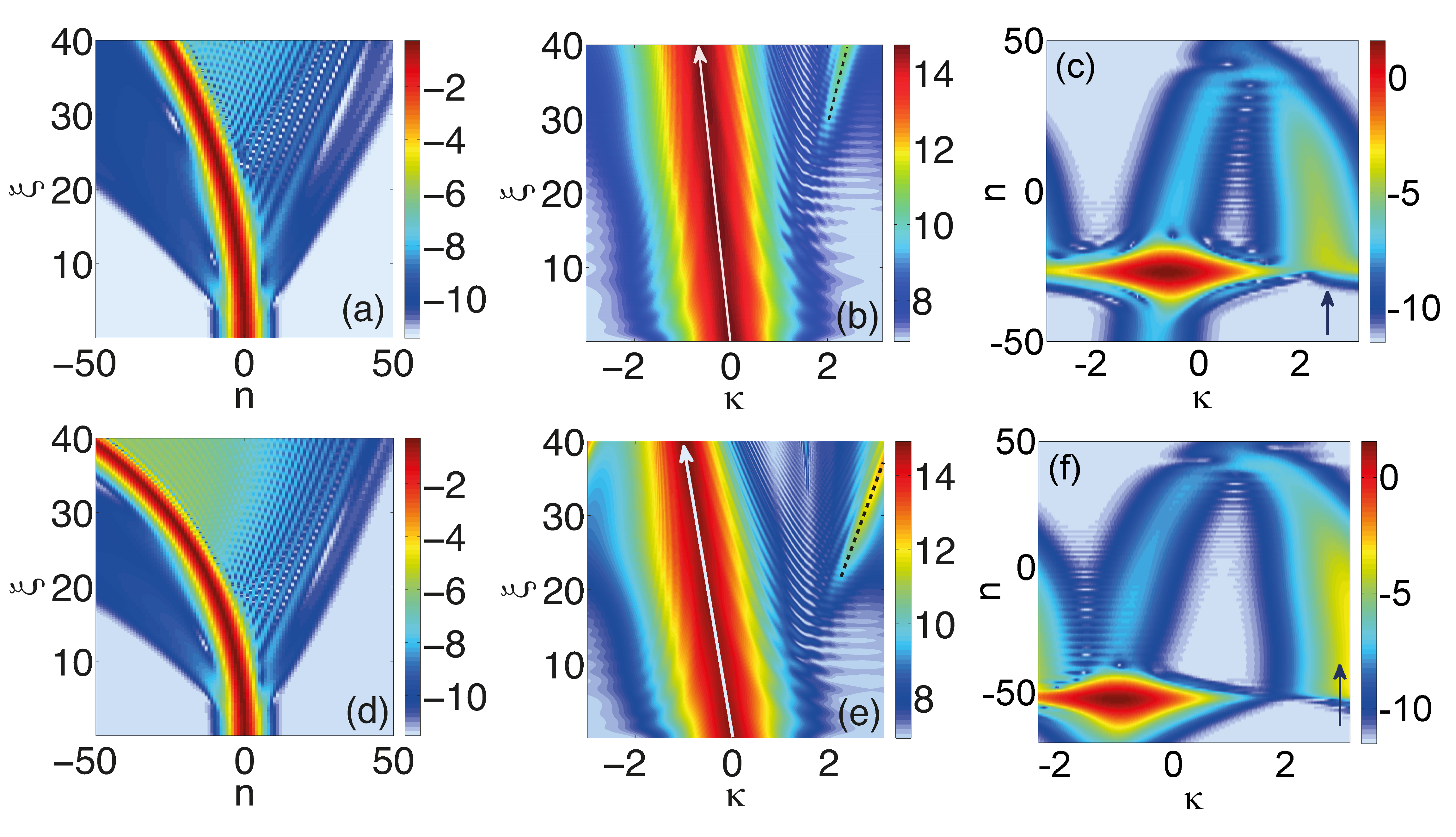}	
				\caption{ Discrete soliton propagation in (a),(d) $(n,\xi)$  and (b), (e) $(\kappa,\xi)$ plane for  $\psi_0=0.8$, $\kappa_0=0$ and $ \delta$=30 nm (for (a),(b)), $ \delta$=50 nm (for (d),(e)) . Dynamic DifRR is evident in $\kappa$-space. (c),(f) XFROG diagram at $\xi=40$ where DifRR is indicated by arrows.}
				\label{fig:cwa_evo}
			\end{center}
		\end{figure}
In Fig.(\ref{fig:cwa_evo}) we demonstrate the  propagation of the discrete soliton for two different values of $\delta$ by numerically solving the governing equation Eq.(\ref{eq:norm_CME}). We observe that for non-zero $\delta$, the soliton changes it wavenumber linearly along its propagation (see Fig.(\ref{fig:cwa_evo}a,d)) and experiences an accelerated motion in the spatial ($n$) domain (see Fig.(\ref{fig:cwa_evo}b,e)) which is quite different from what we observed in uniform WA. The rate of  wavenumber shift and spatial acceleration are increased as the value of $\delta$ is increased. In Fig.(\ref{fig:cwa_evo}b,e) the arrows indicate the linear shift of soliton wave-number $\kappa$ due to the chirping of the waveguide which conceptually introduces a linear potential. From Fig.(\ref{fig:cwa_evo}a,d) we can see that the soliton changes its spatial position from the central waveguide ($n$=0) and leaves behind a plane wave like radiation propagating in the opposite direction. The dynamic nature of DifRR is prominent in ($\kappa,\xi$) plane where it shifts along the propagation distance as indicated by tilted dashed line in Fig.(\ref{fig:cwa_evo}b,e). The intensity of the generated DifRR and its position (at  $\kappa$ plan) can be controlled by $\delta$. For higher values of $\delta$,  stronger DifRR are generated at relatively shorter propagation distance. The XFROG diagrams in Fig.(\ref{fig:cwa_evo}c,f) clearly represent the formation of discrete soliton and DifRR (indicated by arrows). Due to anomalous recoil \cite{19}, a part of the DifRR falls on the other side of the Brillouin boundary when the chirp is strong enough (e.g $\delta > 40$ nm). The variation of the coupling coefficient due to the irregularities in WA can be approximated as an external potential. However an equivalent strength of the potential is difficult to extract from the governing equation (Eq.~\ref{eq:norm_CME}). In an attempt, numerically we try to extract the relationship between the potential strength ($\chi$) and chirp parameter ($g_c=\delta/d_0$). In Fig.(\ref{fig:sol_rr_pos}a) we illustrate the variation of soliton wave-number ($\kappa$) along propagation distance ($\xi$) which clearly shows a linear relationship. The slope of the linear variation depends on $\delta$   (or $g_c$). 

		\begin{figure}[h]
			\begin{center}
				\includegraphics[width=1.7\linewidth]{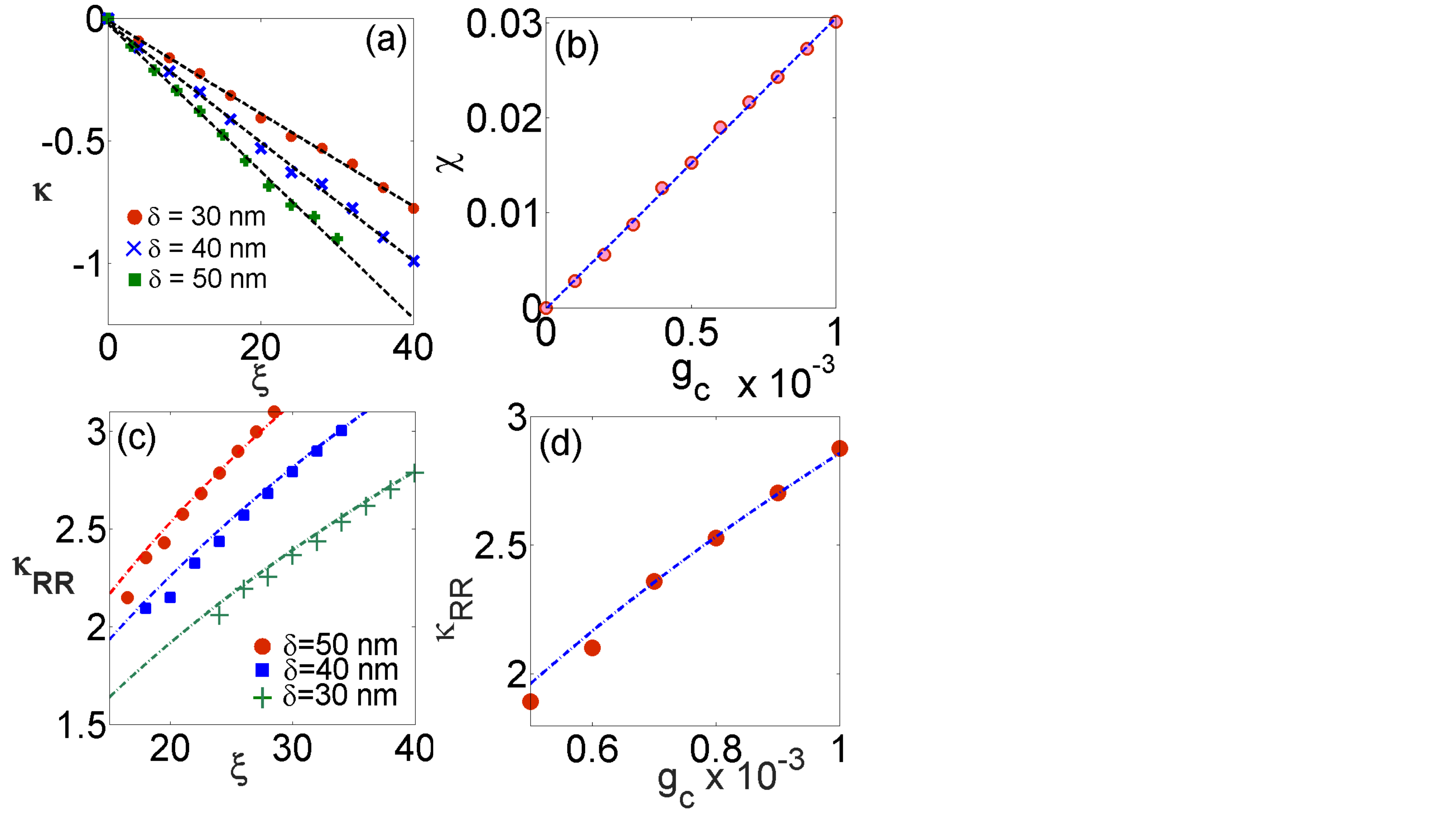}
				\caption{Position of soliton and difRR in the wavenumber domain as a function of propagation distance for different values of $\delta$. (b) Evolution of DifRR for different chirp parameter $\delta$. The dotted lines are obtained using the modified PM relation \ref{eq:mod_PM}}
				\label{fig:sol_rr_pos}
			\end{center}
		\end{figure}
\noindent Inspired by the the Variational results, we can propose an approximate equation to describe the evolution of the soliton wavenumber as,
		\begin{align}
			\kappa(g_c,\xi) = \kappa_0 - \chi(g_c) \xi,
			\label{eq:k_v_xi}
		\end{align}
 where $\chi(g_c)$ is related to effective potential arising due to the chirp parameter $g_c$. The wave number shift ($\Delta \kappa$) of propagating soliton is noted for several $g_c$ which follows a linear relation. Based on the numerical fit as shown in Fig. (\ref{fig:sol_rr_pos}b) we establish an empirical relation between potential strength ($\chi$) and chirp parameter ($g_c$): $\chi(g_c)=m_cg_c$ where the slope is calculated to be $m_c\approx30$. With this information we can take into account the variation of $\kappa$ and its dependency on $g_c$ when the soliton propagates through chirped WA. The change on wave-vector leads to a modification of the existing PM equation Eq.(\ref{PMC}) where we have to impose the linear variation of wave-number as a function of $g_c$ and propagation distance ($\xi$).  The modified PM equation reads, 
\begin{align}
\left[\cos(\kappa_{RR})-\cos(\kappa(g_c,\xi)) + \sin(\kappa(g_c,\xi))\Delta \kappa\right] = \widetilde{\Psi}_0^2
		\label{eq:mod_PM}
		\end{align}
	where, $\Delta\kappa \equiv \kappa_{RR} - \kappa(g_c,\xi)$.
	For a simplified analysis, we consider only the evolution of the soliton wavenumber and compare the position of DifRR obtained from the modified PM equation through Eq.(\ref{eq:mod_PM}) with   numerical results. In Fig.(\ref{fig:sol_rr_pos}c) we demonstrate the evolution of the dynamic DifRR for several $\delta$. The locations of $\kappa_{RR}$ are obtained numerically by solving the governing equation Eq.\eqref{eq:norm_CME} which are in good agreement with the modified PM equation Eq.(\ref{eq:mod_PM}). Finally in Fig.(\ref{fig:sol_rr_pos}c) we demonstrate the variation of $\kappa_{RR}$ with the chirp parameter $g_c$ at a fixed output ($\xi=25$).  The dotted line is obtained from the modified PM expression Eq.(\ref{eq:mod_PM}) which is in good agreement with numerically simulated data (solid dots).

	\subsection{Solitons with an initial non-zero wavenumber ($\kappa_0 \neq 0$)}
In this section, we theoretically and numerically analyze the evolution of a soliton having non-zero initial wavenumber ($\kappa_0 \neq 0$) in linearly chirped WA. The interplay between $\kappa_0$ and chirping parameter brings versatility in the soliton dynamics and allow us to investigate the operating domain never explored before. The wave-vector of the propagating field shifts linearly due to the perturbation imposed by the waveguide chirping. As a consequence, the propagating solitons get self-accelerated which is also theoretically predicted by variational method. For numerical analysis, we consider a soliton propagation for a fixed chirp vale $\delta=$ 30 nm. Here we can have two cases, $\kappa_0 < 0$ and $\kappa_0 >0$. For positive initial wave number ($\kappa>0$) we observe a striking feature where soliton emits twice during its propagation. 

		\begin{figure}[h]
			\begin{center}
							\includegraphics[width=1.25\linewidth]{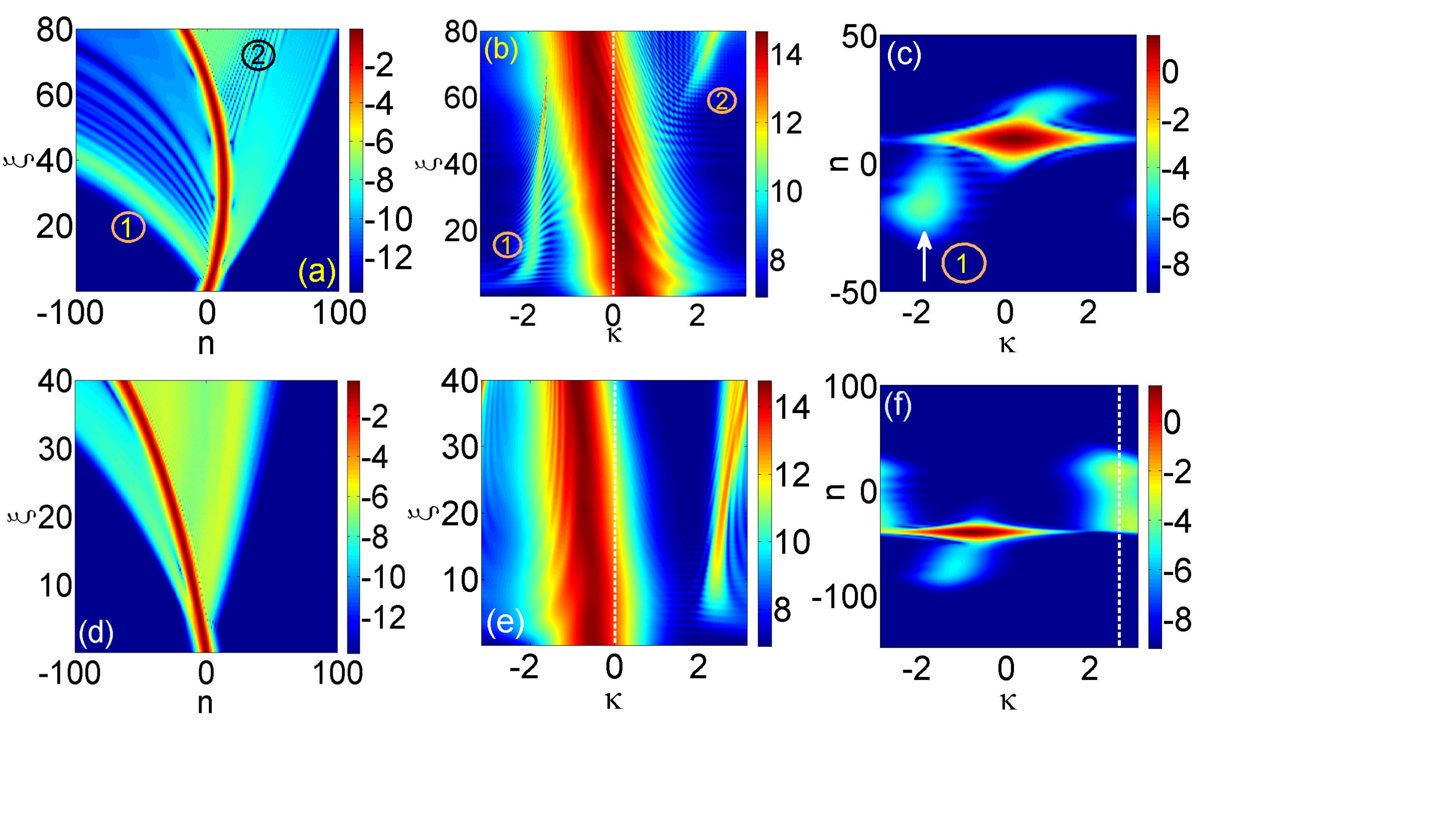}	
				\caption{ Discrete soliton propagation in (a),(d) $(n,\xi)$  and (b), (e) $(\kappa,\xi)$ plane for  $\psi_0=0.8$, $ \delta$=30 nm and $\kappa_0$=0.5 (for (a),(b)), $ \kappa_0$=-0.5 (for (d),(e)) . (b) Dual and (e) single DifRR are evident in $\kappa$-space. (c),(f) XFROG at $\xi=40$ where diffractive radiations are indicated by arrows.}
				\label{Fig7}
			\end{center}

		\end{figure}
In Fig.(\ref{Fig7}a) we illustrate the soliton dynamics in $n$-space where the accelerated soliton emits two consecutive radiations  marked by numbers 1 and 2. In $\kappa$-space, as shown in Fig.(\ref{Fig7}b), the radiations are prominent and showing its dynamic nature. The 1$^{st}$ radiation appears at around $\xi\approx 5$ where as the 2$^{nd}$ begins at $\xi\approx 50$. To the best of our knowledge, the dual DifRR emitted by discrete soliton is never explored before. Fig.(\ref{Fig7}b) helps us to understand qualitatively the possible reason of dual DifRR. The transverse wave-number ($\kappa$) of the propagating soliton shifts linearly with a negative slope due to waveguide chirping. For initial positive wave-number ($\kappa_0>0$) there will be a cross over when $\kappa$ shifts from positive to negative value due to continuous wave-number shift. In Fig.(\ref{Fig7}b) we can observe this cross-over of wave-number which occurs around $\xi \sim 30$. With suitable choice of parameter it is possible that the PM equation (Eq.\eqref{eq:mod_PM}) can be satisfied for $\kappa>0$ as well as $\kappa<0$ which leads to two independent radiations. In Fig.(\ref{Fig7}c) we capture the XFROG of the entire dynamics at a fixed distance $\xi=15$ where the 1$^{st}$ radiation is evident and 2$^{nd}$ radiation is yet to appear.
\begin{figure}[h]
	\begin{center}
		\includegraphics[width=1\linewidth]{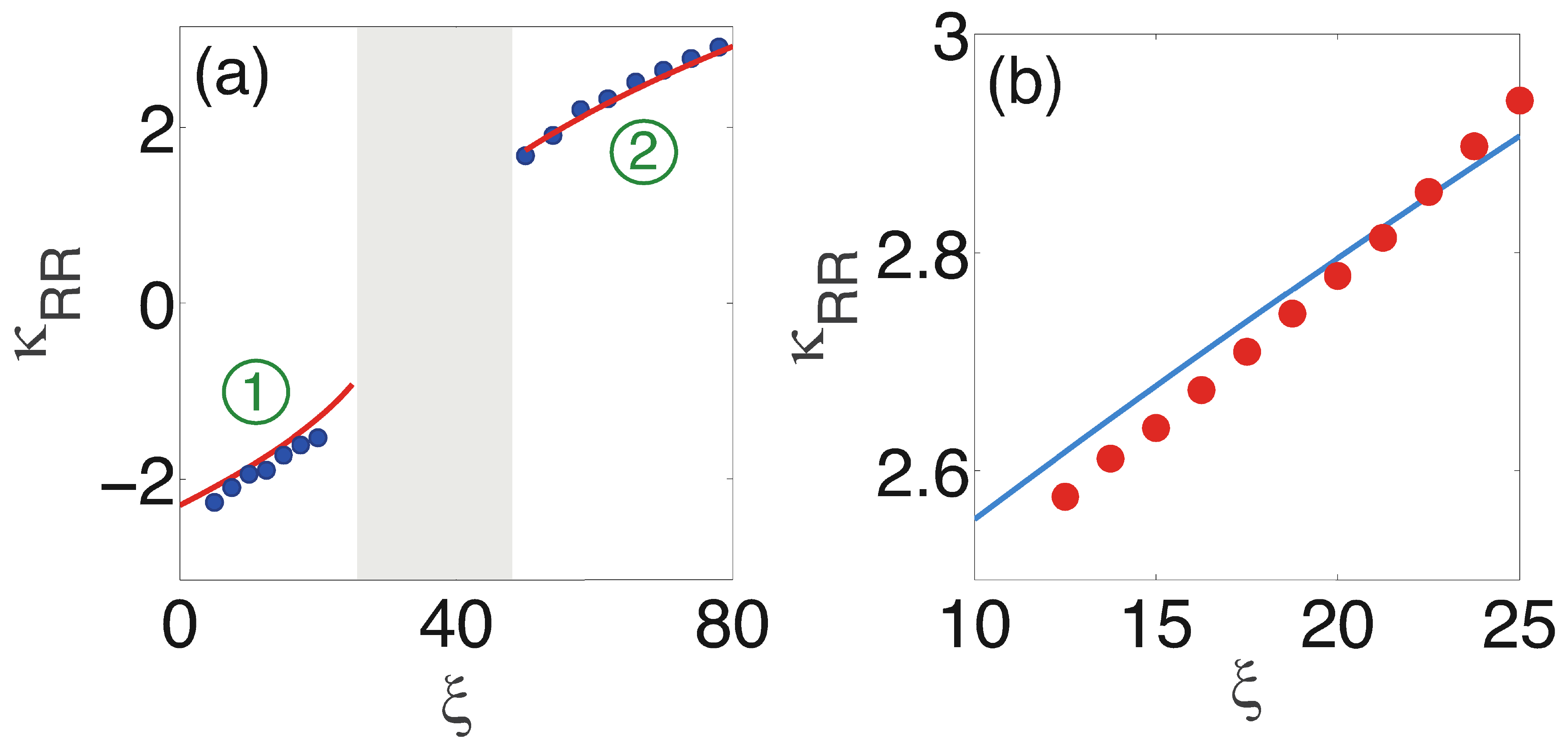}
		\caption{(a) $\kappa_{RR}$ as a function of $\xi$ for dual radiation with the parameters $\kappa_0=0.5$, $\psi_0=0.8$ and $\delta=30$ nm. The two branches indicate two radiations that appear at two different range of distances. (b) $\kappa_{RR}$ as a function of $\xi$ for single radiation for the parameters $\kappa_0=-0.5$, $\psi_0=0.8$ and $\delta=30$ nm. In the figures, the solid lines represent analytical prediction obtained from Eq.\eqref{eq:mod_PM} where dots are numerical data. } 
		\label{Fig8}
	\end{center}
\end{figure}

In Fig.(\ref{Fig7}d-f) we demonstrate the complete dynamics of discrete soliton with negative initial wave-number ($\kappa_0<0$). It is to note that, in case of $\kappa_0<0$, there is no cross-over of soliton wave-number and its value remains negative throughout the propagation. Under such condition only one solution appears from Eq.\eqref{eq:mod_PM} and we observe a single strong radiation. The soliton and the detuned wavenumber of generated DifRR are well separated in the $\kappa$-space exhibiting a dynamic evolution (see Fig.(\ref{Fig7}e)). The generated DifRR is moving away from the soliton owing to the effective linear potential induced by chirp. However due to the boundary of $-\pi$ to $\pi$ set by the one-dimensional lattice, the DifRR emerges from the other side due to anomalous recoil by undergoing a phase shift of $2\pi$. In Fig.(\ref{Fig7}f) we demonstrate the spectrogram where DifRR is evident and indicated by a vertical dotted line. From this figure we also have the hint of anomalous recoil which appears at Brillouin boundary. Finally in Fig.(\ref{Fig8}) we demonstrated the evolution of DifRR theoretically supported by  Eq.\eqref{eq:mod_PM} (solid lines). The dual radiation is evident in Fig.(\ref{Fig8}a) for $\kappa_0>0$ where the soliton emits twice. Two distinct solutions appear when we take into account the cross-over of the wave-number ($\kappa$) in the PM equation (Eq.\eqref{eq:mod_PM}). The shaded region indicates no radiation zone. The solid dots in Fig.(\ref{Fig8}a) represent the  values of $\kappa_{RR}$ extracted from the numerical solution of Eq.\eqref{eq:norm_CME}. In Fig.(\ref{Fig8}b) we depict the case for $\kappa<0$ where a single strong radiation is emitted from moving soliton. The soliton wave-number ($\kappa$) remains negative through-out the propagation and leads to a single solution of Eq.\eqref{eq:mod_PM}. The analytical solution (solid line) corroborates well with the numerical values of $\kappa_{RR}$ indicated by solid dots.

\section{Conclusion}
	
We demonstrate that a linearly chirped waveguide array exhibits discrete-soliton mediated dynamic diffractive resonance radiation where wave-number of the radiation field shift along propagation distance.  Perturbation due to irregularities of waveguide arrangement is modeled as a linear potential in a nonlinear Schr{\"o}dinger  equation which governs the soliton dynamics. We propose realistic waveguide design where diffractive resonance radiation can be excited naturally from a discrete soliton. To gain the intuitive insight of soliton evolution inside a transversely chirped WA we exploit perturbative variational analysis. The variational treatment leads to the equation of motions of soliton parameters which predict self-acceleration of discrete soliton  and linear wave-number shift. Using this information we model the discrete nonlinear Schr{\"o}dinger equation and theoretically modify the phase-matching equation which capture the dynamic nature of wave-number shift of diffractive radiation. Evolution of the soliton is investigated for zero and non-zero initial wave-number. An intriguing effect of dual diffractive resonance radiation is observed for the first time when a soliton with positive wave-number is launched in the WA. The theoretical underpinning of dual radiation lies with the fact that, the  soliton  wave number experiences a cross-over by shifting its value from positive to negative owing to the chirping in WA. This cross-over results two distinct solution of PM equation at two different propagation distances and leads to dual radiation. We theoretically confirm this phenomenon by solving the modified PM equation. This work could pave the way
for designing waveguide-array based optical devices that are capable of generating controllable spacial supercontinuum.

\begin{acknowledgments}
A.P.L acknowledges University Grants Commission (UGC), India for support through a research fellowship.
\end{acknowledgments}

\end{document}